\documentclass[10pt,twocolumn]{article}
\usepackage{amsmath}
\usepackage{amssymb}
\usepackage{epsfig}

\begin{document}

\setlength{\baselineskip}{12pt}

\title{Clustering Algorithms for Scale-free Networks and
Applications to Cloud Resource Management}
\author{Ashkan Paya and Dan C. Marinescu \\
Computer Science Division \\
Department of Electrical Engineering and Computer Science \\
University of Central Florida, Orlando, FL 32816, USA \\
Email:ashkan\_paya@knights.ucf.edu, dcm@cs.ucf.edu}

\maketitle

%\tableofcontents

\begin{abstract}
In this paper we introduce algorithms for the construction of scale-free networks and for clustering around the {\it nerve centers}, nodes with a high connectivity in a scale-free networks. We argue that such overlay networks could support self-organization in a complex system like a cloud computing infrastructure and allow the implementation of optimal resource management policies.
\end{abstract}

\section{Introduction and Motivation}

The analysis of high-level models of a system allow us to better understand its behavior. Oftentimes we use a finite state machine model of a system where vertices represent states and the directed arcs represent transitions between states. Such models provide insights on the system dynamics, but are seldom used for the analysis of complex systems. Complex systems have a very large state space and there are many possible transitions between states.

The possible interactions of the entities in complex
biological, social, economic, or computing system can also be described by a graph where vertices represent active entities and the edges represent the communication channels between them. Though these models provide only static information, they capture some important properties of the system and can be very useful to decide if a system is scalable, in other words if the system organization is capable to accommodate growth. In this paper we argue that such a graph could also reveal if the internal organization of a system is compatible with self-organization and self-management principles.

There are strong arguments supporting the belief that self-organization and self-management are highly desirable for dynamic, large-scale systems. In such cases, a centralized or even a traditional distributed decision making processes cannot ensure an optimal system behavior. The very large volume of state information, the rapid pace of state changes of individual components, and  the long communication delays require a different approach for system management and control.

The alterative we discuss in this paper is to allow individual entities to make decisions based primarily on local information. Yet, all entities must cooperate to implement the system objectives and policies thus, some form of coordination among the entities is necessary. To satisfy these contradictory requirements, a relatively small subset of entities must act as the {\it nerve} centers of the system and perform control functions, while the other entities carry out actions as directed by these nerve centers.  We shall call {\it core} the entities performing control functions and {\it server} the others  entities. This terminology is justified by the application to self-management in cloud computing discussed in Section \ref{ComputerClouds}.

The approach we propose should allow each entity to decide to which one of the two classes, core or server it belongs to, based on intrinsic properties of the entity,  including its state. Once this decision is  made, a server should be able to join the cluster built around one of the core entities. This decision should be based on some distance metric; then the server should work in concert with the other servers in the cluster, following the directives of the core entity which assumes a control role for the cluster.

To ensure system agility and allow the system to promptly react to rapid state changes,  the amount of state information maintained by a server should be minimal; the entity should only be aware of its immediate neighbors and of the core entity leading the cluster it has decided to join. A core entity should be aware of all the members of its cluster and of a subset of the other  core entities. This strategy could reduce the total amount of state information, provided that: (i) the core entities are well connected and (ii) most of the servers have only a few connections.

At the same time, the core and the server entities should be able to communicate efficiently. A core entity should be able to monitor the status of the servers in its cluster and disseminate policy-related information. In turn, a server should be able to initiate communication with the core whenever necessary. Such an organization requires the development of a {\it virtual} communication infrastructure  superimposed on the {\it physical} communication infrastructure. A  {\it scale-free} network supports self-organization and ensures that: (i) the system is scalable; (ii) the system organization can be done only based on local information; and (iii) resource management decisions can be made based on local, more accurate state information, rather than global state information.

\section{Scale-free organization}
\label{ScaleFreeNetworks}

Many complex systems enjoy a {\it scale-free organization} \cite{Barabasi99,Barabasi00}.  In a scale-free
organization the probability $p(k)$ that an entity interacts with $k$ other entities decays as a power law

\begin{equation}
p(k) \approx k^{-\gamma}
\end{equation}
with $\gamma$ a constant and $k$ a positive integer. This probability is independent of the type and the function of the system, the identity of its constituents, and the relationships between them.

Empirical data for many man-made systems confirm the existence of scale-free networks. Examples abound,
e.g., the power grid of the Western US has some $5,000$ vertices representing
power generating stations; in this scale-free network $\gamma \approx 4$. The scale-free organization appears naturally in social networks. For example, the collaborative graph of movie actors where links are present if two actors were ever cast in the same movie follows the power law with $\gamma \approx 2.3$.  The probability that $q$ pages of the World Wide Web  point to one page is $p(k) \approx k^{-2.1}$ \cite{Barabasi00}. Recent studies indicate that $\gamma \approx 3$ for the citation of scientific papers. The larger the network, the closer a power law with $\gamma \approx 3$ approximates the distribution \cite{Barabasi99}.

Several models of graphs have been investigated starting with the Erd\"os-R\'eny model \cite{Erdos59} where the number of vertices is fixed and the edges connecting vertices are created randomly. This model produces a homogeneous network with an exponential tail; connectivity follows a Poisson distribution peaked at the the average degree $\bar{k}$ and decaying exponentially for $k >> \bar{k}$. An evolving network, where the number of vertices increases linearly and a newly introduced vertex is connected to $m$ existing vertices according to a preferential attachment rule is described by Barab\'asi and Albert in \cite{Albert99,Albert00,Barabasi99}.

Regular graphs where a fraction of edges are rewired with a probability $p$ have been proposed by Watts and Strogatz and called small-worlds networks \cite{Watts98}. Networks whose degree distribution follows a power law are called scale-free networks. The four models are sometimes referred as ER (Erd\"os-R\'eny), BA (Barab\'asi - Albert), WS (Watts-Strogatz), and SF (Scale-free) models, respectively \cite{Goh01}.

Throughout this paper we shall use the terms networks, nodes, and links when we discuss a physical system; we shall use the terms graphs, vertices, and arcs when we discuss the model of a system.

The degree distribution of scale-free networks follows a power law; we only consider the discrete case when the probability density function is  $p(k) = a f(k)$  with $f(k) = k^{-\gamma}$ and the constant $a$ is $a= 1/\zeta(\gamma, k_{min})$ thus,

\begin{equation}
\label{17}
p(k)= { 1 \over {\zeta(\gamma, k_{min})}}k^{-\gamma}.
\end{equation}
In this expression $k_{min}$ is the lowest degree of any node, and for the applications we discuss in this grant request $k_{min}=1$; $\zeta$ is the Hurvitz zeta function\footnote{The Hurvitz zeta function  $\zeta(s,q)= \sum_{n=0}^{\infty} { 1 \over {(q+n)^{s}} }$ for $s,q \in \mathbb{C}$ and $\mathfrak{Re}(s) > 1$ and $\mathfrak{Re}(q) > 0$. The Riemann zeta function is $\zeta(s,1)$.}

\begin{equation}
\zeta(\gamma, k_{min}) = \sum_{n=0}^{\infty} {1 \over {(k_{min}+n)}^{\gamma}} =
\sum_{n=0}^{\infty} {1 \over {(1+n)}^{\gamma}}.
\end{equation}

\begin{figure}[!ht]
\begin{center}
\includegraphics[width=7cm]{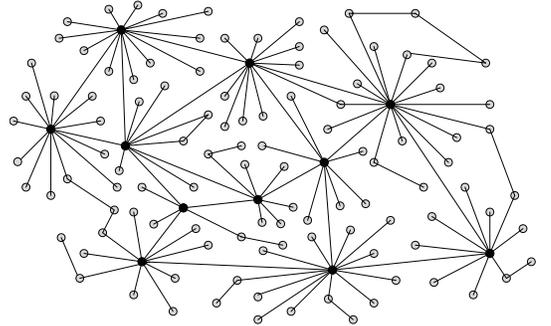}
\end{center}
\caption{A scale-free network is non-homogeneous; the majority of
the vertices have a low degree and only a few vertices are connected to a
large number of edges; the majority of the vertices are directly connected with the vertices with the highest degree.}
\label{ScaleFreeNetworkFig}
\end{figure}

A scale-free network is {\it non-homogeneous}; the majority of the nodes have a low degree and only a few nodes are connected to a large number of links, Figure \ref{ScaleFreeNetworkFig}. The average distance $d$ between the $N$ nodes, also referred to as the diameter of the scale-free network, scales as $\ln N$; in fact it has been shown that when $k_{min} > 2$ a lower bound on the diameter of a network with $2 < \gamma < 3$ is $\ln \ln N$ \cite{Cohen03}.

A number of studies have shown that scale-free networks have remarkable properties such as: (a) robustness against random failures \cite{Barabasi00}; (b) favorable scaling \cite{Albert99,Albert00}; (c) resilience to congestion \cite{Goh01}; (d) tolerance to attacks \cite{Toroczkai04}; and (e) small diameter \cite{Cohen03} and small average path length \cite{Barabasi99}.
The moments of a power law distribution play an important role in the behavior of a network. It has been shown that the giant connected component (GCC) of networks with a finite average vertex degree  and divergent variance can only be destroyed if all vertices are removed; thus, such networks are highly resilient against faulty constituents \cite{Mondal07}.

These properties make scale-free networks very attractive for interconnection networks in many applications including social systems \cite{Newman01}, peer-to-peer systems, sensor networks \cite{Marinescu10} and, as we argue in this paper, cloud computing.

Another important property is that the majority of the nodes of a scale-free network are directly connected with the nodes of higher degree, see Figure \ref{ScaleFreeNetworkFig}. For example, in a network with $N=130$ nodes and $m=215$ links $60\%$ of the nodes are directly connected with the five nodes with the highest degree, while in a random network fewer than half, $27\%$, have this property  \cite{Albert00}. Thus, the nodes of a scale-free network with a degree larger than a given threshold $T$ could assume the role of ``core nodes'' and assume management functions; the other nodes assume the role of computational and storage servers. This partition is autonomic; moreover, most of the server nodes are at distance one, two, or three from a core node which could gather more accurate state information from these nodes and with minimal overhead. In the next example if $k_{lim}=4$ then $92.5\%$ of the nodes are servers.

As an example, consider the case $\gamma=2.5$ and the minimum node degree, $x_{min}=1$; we first determine the value of the zeta function  $\zeta(\gamma,x_{min})$ and  approximate $\zeta(2.5,1)=1.341$ thus, the distribution function is
$p(k)= k^{-2.5}/1.341 = 0.745 \times( 1 / k^{2.5})$, where $k$ is the degree of each node. The probability of nodes of  degree $k > 10$ is $\text{Prob}(k > 10) = 1 - \text{Prob} (k \le 10) = 0.015$. This means that at most $1.5\%$ of the total number of nodes will have more than $10$ links connected to them; we also see that $92.5\%$ of the nodes have degree $1,2$ or $3$. Table \ref{NumberOfComponents} shows the number of nodes of degrees $1$ to $10$ for a very large network, $N=10^{8}$.

\begin{table}[ht!]
\begin{center}
\caption{A power-law distribution with degree $\gamma=2.5$; the probability, $p(k)$, and $N_{k}$, the number of nodes with degree $k$, when the total number of vertices is $N=10^{8}$.}
\

\label{NumberOfComponents}
\begin{tabular} {|ccc||ccc|}
\hline
 $k$ & $p(k)$ & $N_{k}$ & $k$ & $p(k)$ & $N_{k}$ \\
 \hline
 1 & 0.745 & $ 74.5 \times 10^{6}$ &  6 & 0.009 & $ 0.9 \times 10^{6}$ \\
 2 & 0.131 & $ 13.1 \times 10^{6}$ &  7 & 0.006 & $ 0.6 \times 10^{6}$ \\
 3 & 0.049 & $ 4.9 \times 10^{6}$  &  8 & 0.004 & $ 0.4 \times 10^{6}$ \\
 4 & 0.023 & $ 2.3 \times 10^{6}$  & 9 & 0.003 & $ 0.3 \times 10^{6}$  \\
 5 & 0.013 & $ 1.3 \times 10^{6}$  & 10 & 0.002 & $ 0.2 \times 10^{6}$ \\
 \hline
\end{tabular}
\end{center}
\end{table}

\section{Centralized Clustering Algorithm}
\label{CentralizedAlgorithm}

When the number of edges of the graph (or, equivalently, the number of nodes of a physical network) is relatively small, $N \le 10,000$, the creation of a scale-free network  and then clustering can be carried out in a centralized manner. In this case a master has information about all the nodes and runs the algorithm discussed in this section to first construct a scale-free network and then to split the set of nodes into clusters.

Clustering in a scale-free network is the process of creating groups of server nodes around each core node. Each server node is assigned by the central authority to the cluster built around the core node to minimize the distance between the two; when a server node is equally distant from several core nodes, then it is assigned randomly to one of them.

\bigskip

{\bf The input.} We start with a network modeled as a fully connected graph, rather than a random graph.  Our goal is to rewire this network as a scale-free one.

The algorithm assumes a known number of nodes, $N$, and a given exponent $\gamma$  of the degree distribution, $2 \le \gamma \le 3$. We also assume that the individual nodes are uniquely identified by integers, $nId \in [1,N]$. Another parameter of the algorithm is the threshold $T$, a positive integer used to separate core from server nodes; nodes of degree larger or equal to $T$ are core nodes, the other are server nodes.

The number $T$ of rewiring iterations cannot be predicted due to the randomness of the algorithm. Two stopping criteria for the number of iterations are possible:

\begin{enumerate}
\item
After several iterations we compute the distance between the desired degree distribution, a power law distribution with the exponent $\gamma$, and the current degree distribution. We stop when this distance is smaller than $\epsilon$, a small constant which captures our desired accuracy of the algorithm. We use the methodology described later in this section to compute the distance between two distributions of a discrete random variables.
\item
Fixed (empirically determined) number of iterations.
\end{enumerate}

\bigskip

{\bf The algorithm.} The algorithm has two phases: (1) rewiring of the physical network to create a scale-free organization; (2) clustering. In this process the degree of the node $nId=i$ is denoted as $Deg(i)$, a component of the $N$-dimensional vector $Deg$. The $Links$ vector contains information about the links of the scale-free network, $Links(l)=(i,j)$ means that the $l$-th link connects nodes $i$ and $j$.

\medskip

\underline{Phase 1.} Calculate the parameter $\alpha$ as

\begin{equation}
\alpha={\frac{1}{1-\gamma}}.
\end{equation}

Initialize the degree of each node and the $Links$ table

\begin{equation}
Deg(i) = 0,~~~ 1 \le i \le N~~~~~\text{and}~~~~~Links(l) = (0,0).
\end{equation}

The iterative graph rewiring process consists of the following steps:

\begin{enumerate}

\item
Pick up randomly a node with $nId=i$ and compute the probability

\begin{equation}
p_{i} = { i^{-\alpha} \over {\sum_{m=0}^{i} m^{-\alpha}}}
\end{equation}

\item
Pick up randomly a node with $nId=j$ and compute the probability

\begin{equation}
p_{j} = { i^{-\alpha} \over {\sum_{m=0}^{j} m^{-\alpha}}}
\end{equation}

\item
Generate a random number $\kappa \in [0,1]$.

\item
Decide that the link connecting nodes $i$ and $j$ can be included in the scale-free network if the following condition is satisfied:

\begin{equation}
 (1 - {e^{-2 N * p_{i} * p_{j}}}) > \kappa.
\end{equation}

\item
If this condition is satisfied record the presence of this link

\begin{equation}
Link(l) = (i,j).
\end{equation}

\item
Increment the degrees of the two terminal nodes of this link

\begin{equation}
Deg(i) = Deg(i) + 1 ~~~~~\text{and}~~~~~ Deg(j) = Deg(j) + 1.
\end{equation}

\item
Check the stoping condition.

\item
If stoping condition is not satisfied execute the next iteration. Else
set $L = \dim{Links}$, the number of links, and go to the next step, the clustering.

\end{enumerate}

\medskip

\underline{Phase 2.} The clustering phase uses:
 \begin{itemize}
 \item
 A two-dimensional $M x (N-M)$ array  $Dist$ with $M$ the number of core nodes and $(N-M)$ the number of server nodes.
 \item
 An array $ClusterId$ with $(N-M)$ entries of the form $(sId, cId)$, where $sId\in [1, N-M]$ is the id of a server node and $cId \in [1,M]$ is the id of a core node.
 \item
 The $Cluster$ array with $M$ entries of the form $cId, (s_{1},...s_{q})$, where $(s_{1},...s_{q})$ are server nodes in the cluster build around the core node $cId$.
 \end{itemize}

The algorithm consists of the following steps:

\begin{enumerate}
\item
Sort the $Deg$ vector and identify the core nodes as

\begin{equation}
Core(k) = i ~~\text{if}~~Deg(i) \ge T.
\end{equation}

Call $M$ the number of core nodes equal to the number of clusters.

\item
Initialize the distance array; $Dist(i,j)=-1$ means that the distance
between nodes $i$ and $j$ has not been computed yet.

\begin{equation}
Dist(m,n) = -1.
\end{equation}

\item
Set $k=1$ and start iterations on core nodes:

\begin{enumerate}
\item
Pick up a core node as a head of the cluster:
\begin{equation}
ClusterHead = Core(k)
\end{equation}

\item
Start iterations to determine the distance of server nodes to the $ClusterHead$.
  \begin{enumerate}
  \item
  Search the $Links$ vector of dimension $L$ for all entries where either the first or the last node is the $ClusterHead$.
  \begin{multline}
  \text{if~~} Links(l) = (i,ClusterHead)\\
~~\text{then}~~Dist(ClusterHead,i)=1.
  \end{multline}
  \begin{multline}
  \text{if~~} Links(l) = (ClusterHead,j)\\
~~\text{then}~~Dist(ClusterHead,j)=1.
  \end{multline}
  \item
  Search the $Links$ vector for all entries where either the first or the last node is a node $d_{1}$ such that $Dist(d_{1},ClusterHead)=1$.
  \begin{multline}
  \text{if~~} Links(l) = (d_{1},j)\\
~~\text{then}~~Dist(ClusterHead,j)=2.
  \end{multline}
  \begin{multline}
  \text{if~~} Links(l) = (i,d_{1})\\
~~\text{then}~~Dist(ClusterHead,i)=2.
  \end{multline}
  \item
  Search the $Links$ vector for all entries where either the first or the last node is a node $d_{2}$ such that $Dist(d_{2},ClusterHead)=2$.
  \begin{multline}
  \text{if~~} Links(l) = (d_{2},j)\\
~~\text{then}~~Dist(ClusterHead,j)=3.
  \end{multline}
  \begin{multline}
  \text{if~~} Links(l) =(i,d_{2})\\
~~\text{then}~~Dist(ClusterHead,i)=3.
  \end{multline}
  \item
  Repeat the process for distances up to a $d_{max}$
  \end{enumerate}
  \item
  Set $k=k+1$
  If $k \le M$ execute the next iteration on the core nodes
\end{enumerate}
 \item
 Determine the cluster each server node should be assigned to.
 \begin{enumerate}
  \item
  Set $iter=1$.
  \item
   Compute the minimum distance of server node $iter$ to all core nodes.
    \begin{equation}
     d_{min} = \min{Dist(iter, cId)}
    \end{equation}
  \item
  Assign the server node to the cluster around a core node at minimum distance
  \begin{equation}
  ClusterId(iter) = cId.
  \end{equation}
  \item
  Set $iter=iter+1$.
  \item
  If $iter \le N-M$ execute the next iteration.
  \item
  Else execute the next step.
   \end{enumerate}
  \item
  Call $Cluster(i),~i \in [1, M]$, the data structure containing the information about the membership in each cluster. This data structure is constructed using the information from $ClusterId(j),~~j \in [1, (M-N)]$.
\end{enumerate}

\bigskip

\begin{figure}[!ht]
\begin{center}
\includegraphics[width=7cm]{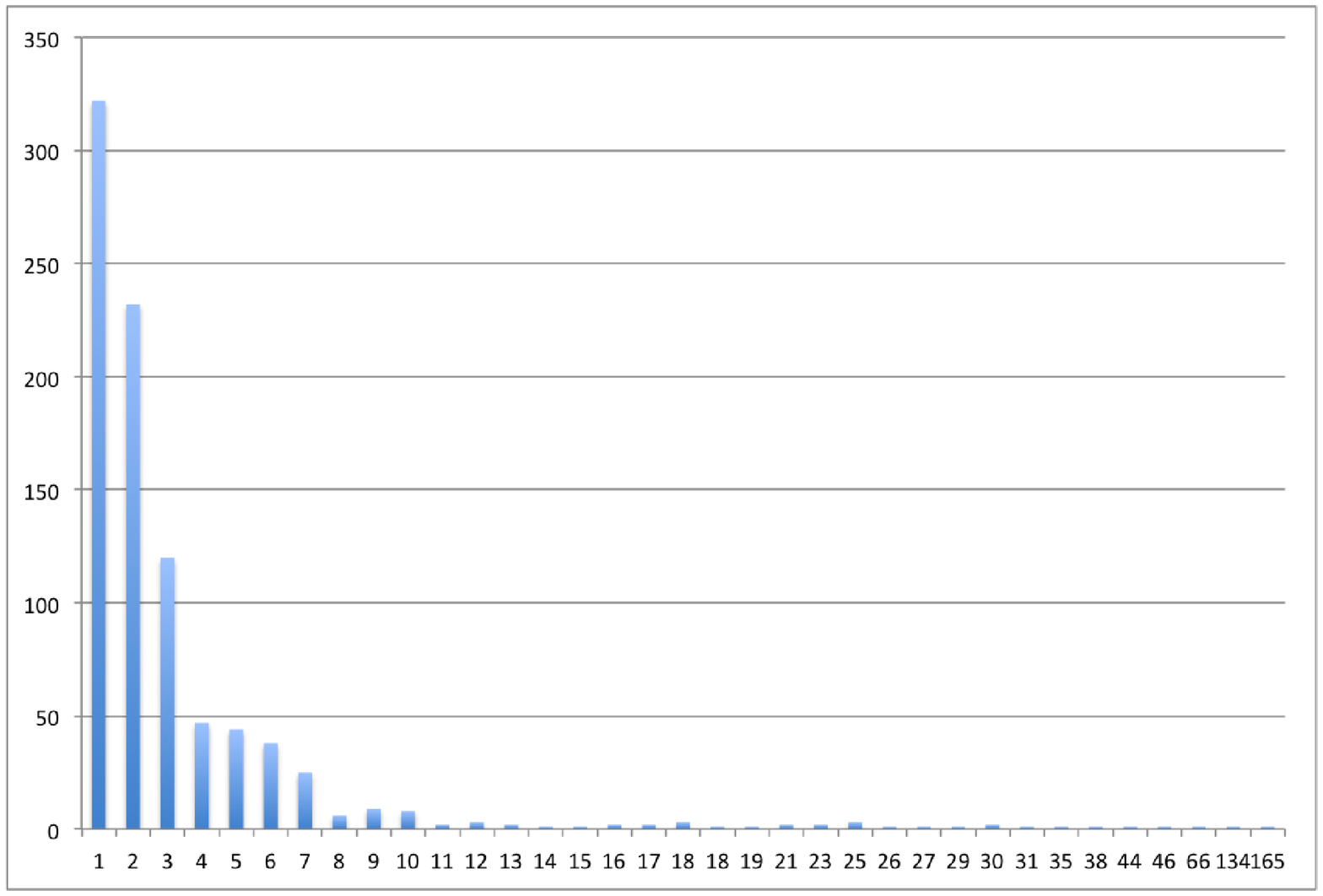}
\includegraphics[width=7cm]{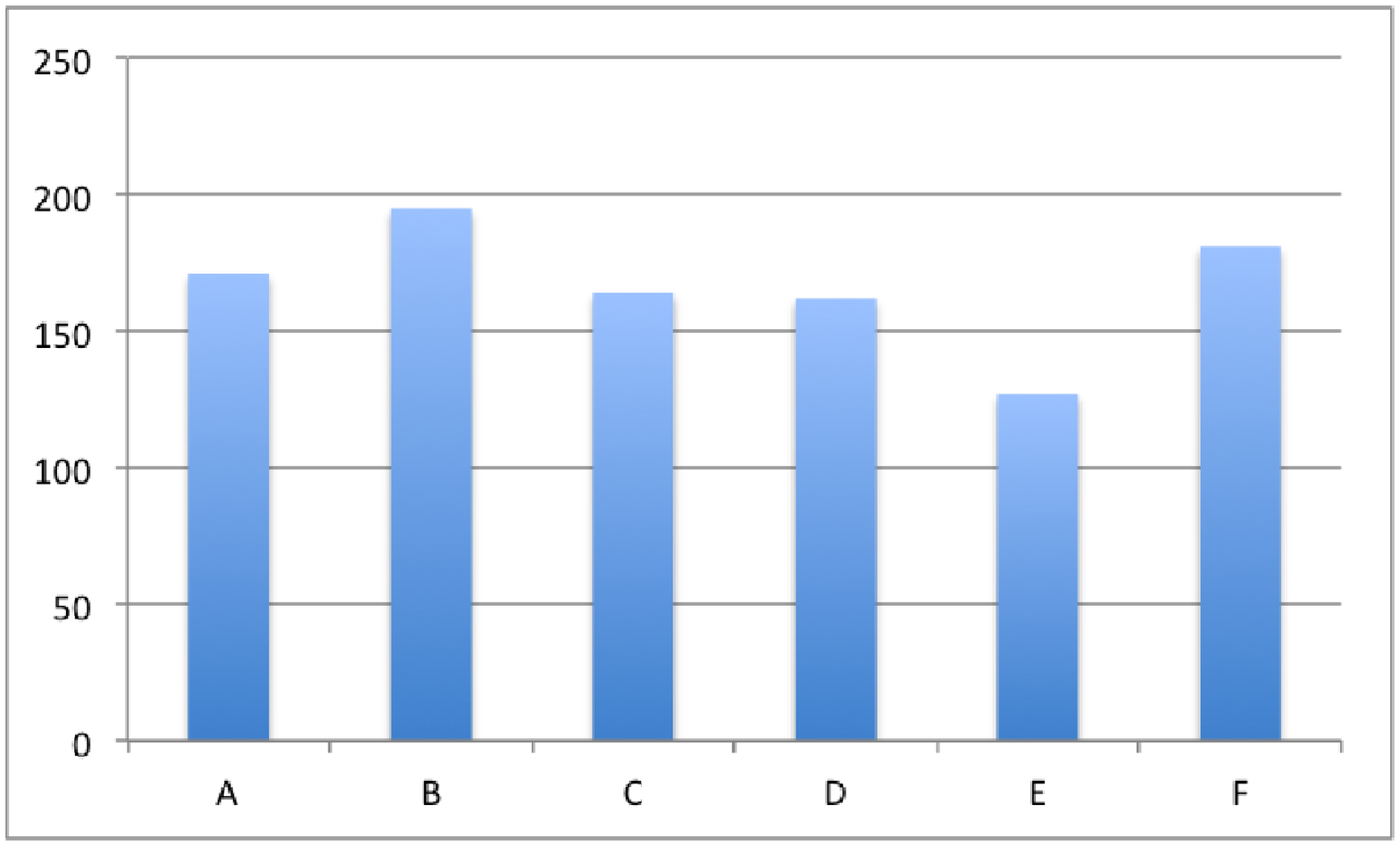}

\end{center}
\caption{The number of nodes $N=1,000$ and $\gamma=2.5$. The number of rewiring iterations is fixed at $1.4 \times N$. The threshold for separation of core and server nodes is $T=32$ and number of core nodes is $M=6$. (Top) The histogram of degree distribution. (Bottom) The  number of server nodes in each one of the six clusters, $A, B, C, D, E$ and $F$.}
\label{C-HistogramFig}
\end{figure}

\begin{figure}[!ht]
\begin{center}
\includegraphics[width=7cm]{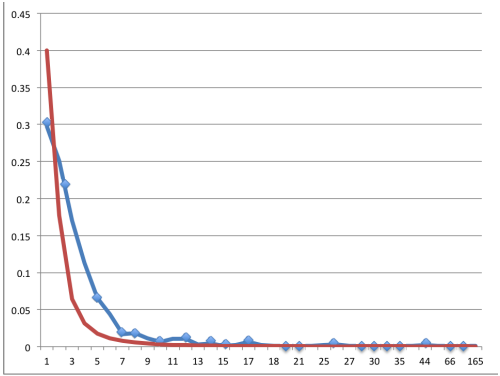}
\end{center}
\caption{The theoretical degree distribution (continuous line) and the degree distribution obtained with the algorithm presented in this section (blue line with dots) for the construction of a scale-free network.}
\label{C-GraphFig}
\end{figure}

{\bf Implementation and results.} The algorithm was implemented in Java. Even though
in our experiments we considered only a relatively small number of nodes, $N=1,000$
we had to resort to several programming tricks to overcome the limitations of the Java heap space.

We choose $\gamma=2.5$ and set the threshold for the separation of core and server node as $T=32$. Under these condition the number of core nodes was $M=6$. We use a stoping condition of the first type, the rewiring stops after $1.4 \times N$ iterations and in this case the error computed as the trace distance is $e \approx 0.2005$. Figure \ref{C-HistogramFig}(a) and (b) show a histogram of the degrees of the nodes and the number of server nodes in each one of the six cluster labeled as $A, B, C, D, E$ and $F$, respectively. Figure \ref{C-GraphFig} plots the theoretical and the  degree distribution obtained using the algorithm  in this section.

\begin{figure}[!ht]
\begin{center}
\includegraphics[width=7cm]{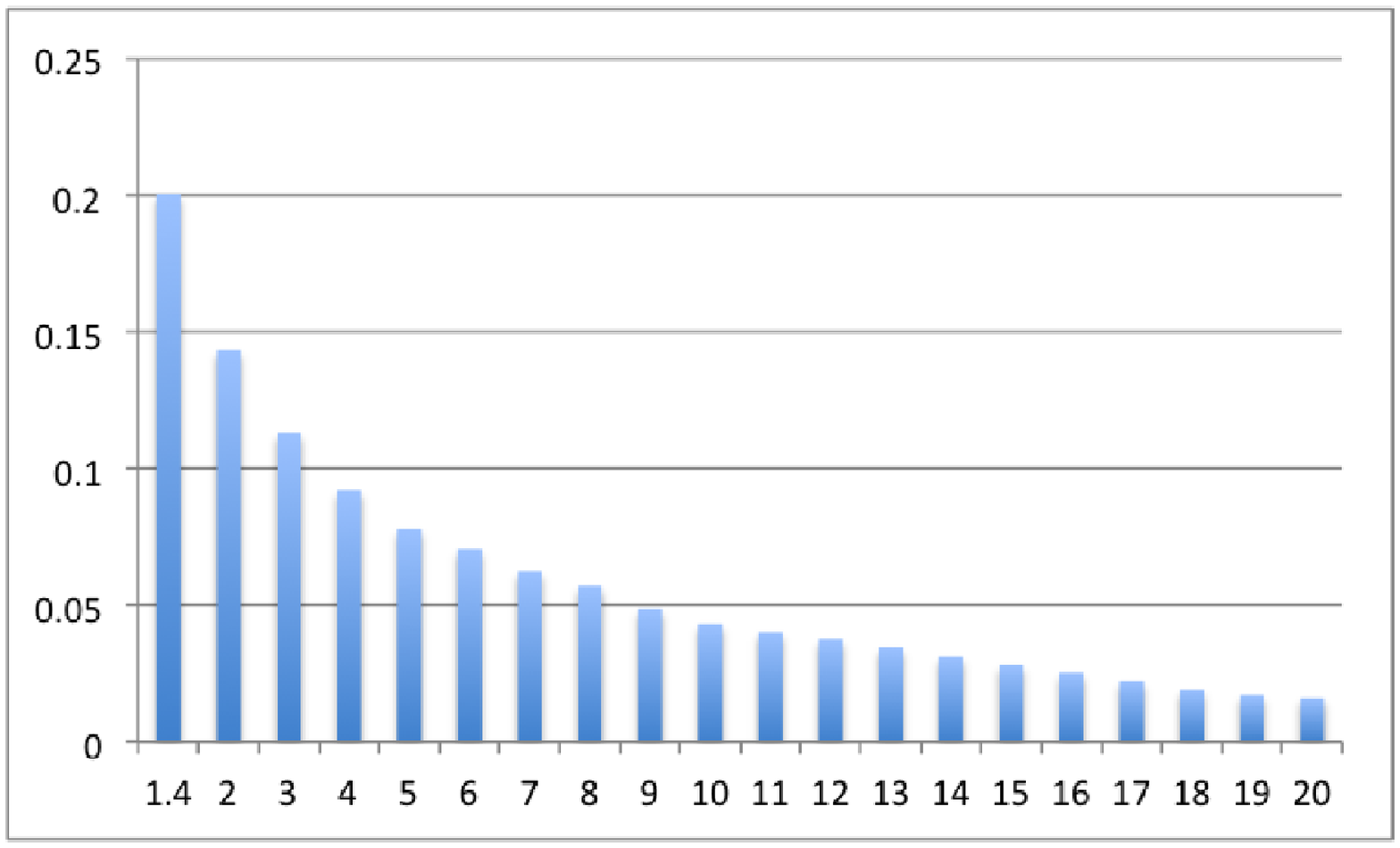}
\includegraphics[width=7cm]{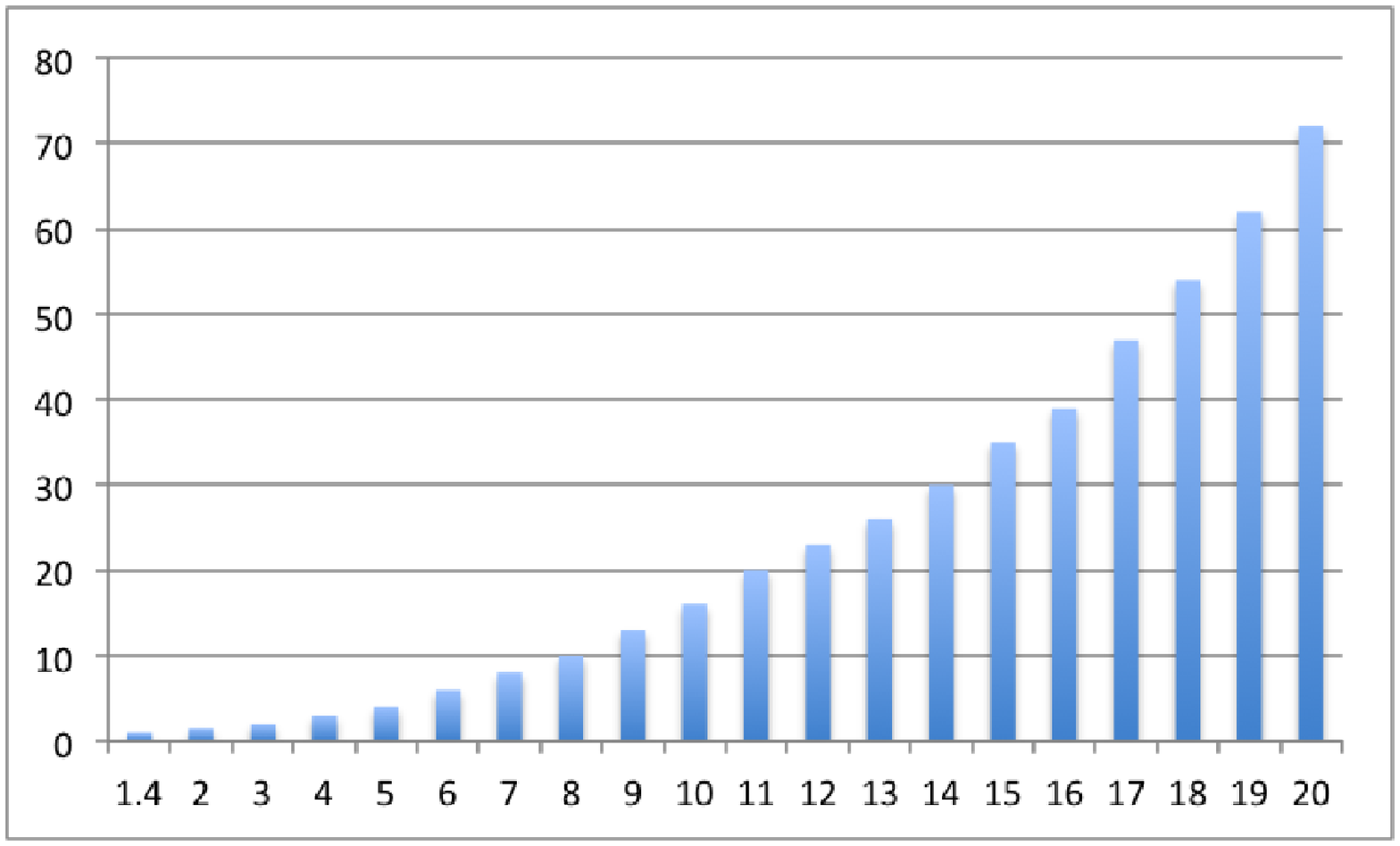}

\end{center}
\caption{The distance between the theoretical and the experimental degree distribution and the execution times function of the number of rewiring iterations. The number of iterations is expressed as multiples of $N$, the number of nodes. (Top) The distance. (Bottom) The execution time in seconds.}
\label{C-IterationsFig}
\end{figure}

One of the advantages of the centralized algorithm is that we can obtain a degree distribution as close to the theoretical one as desired, by increasing the number of rewiring iterations. Figure \ref{C-IterationsFig}(a) and (b) show distance between the theoretical and the experimental degree distribution and the execution times function of the number of rewiring iterations. respectively. We see that the accuracy and the execution time increase exponentially with the number of rewiring iterations.

The execution was done on the Amazon cloud; we used one {\it medium} instance; the execution times on the cloud were comparable with the times when execution was done locally on a system with similar resources as the
ones provided by the AWS (Amazon Web Services) instance.

\bigskip

{\bf The distance between two probability density functions.} Several measures exist for the similarity/dissimilarity of two probability density functions of discrete random variables including the trace distance, fidelity, mutual information, and relative entropy \cite{Kolmogorov65}.
The {\it trace distance} (also called Kolmogorov or L1 distance) of two probability density functions, $p_{X}(x)$ and $p_{Y}(y)$, and their {\it fidelity} are defined, respectively, as
                                                                                                                                          \begin{multline}
\label{28}
D \left( p_{X}(x), p_{Y}(x) \right) = {1 \over 2} \sum_{x} \mid p_{X}(x)- p_{Y}(x) \mid\\
~~~\text{and}~~~
 F \left( p_{X}(x), p_{Y}(x) \right) = \sum_{x} \sqrt{ p_{X}(x) p_{Y}(x)}.
\end{multline}

The trace distance is a metric: it is easy to prove non-negativity, symmetry, the identity of indiscernible, and the triangle inequality. On the other hand, the fidelity is not a metric, as it fails to satisfy the identity of indiscernible,
$F \left( p_{X}(x), p_{X}(x) \right) = \sum_{x} \sqrt{ p_{X}(x) p_{X}(x)}= 1 \ne 0$.

In our experiments we computed the trace distance.

\bigskip

{\bf Observations.} Our choice of a fully-connected graph as the starting model for the construction of the scale-free network is motivated by the application of clustering to cloud resource management discussed in Section \ref{ComputerClouds} when the interconnection network can be modeled as a fully-connected graph. Indeed, in a data center multiple servers mounted on a rack are connected to a GigabitEthernet, an InfiniBand, or a Mirynet switch. The switch is connected to other switches and the  physical network allows all servers to communicate directly with one another.

\section{Algorithms for self-organization}
\label{DistributedAlgorithms}

When the network size is so large that no single site can hold information about the entire network, the construction of the scale-free network and then the clustering can be done by a biased random walk. In a self-organizing system the only information available to an entity is the set of its neighbors; biased random walks allow scale-free organization and clustering in a system where only local information is available to each entity.

%************************************
\subsection{Biased random walks}
\label{BiasedRandomWalks}
%***************************************

We know that many network models have good expansion  properties thus, the second eigenvalue  of their transition matrix is bounded away from one. In this case samples taken from consecutive steps of a random walk can achieve statistical properties similar to independent sampling.

Unfortunately, the application of random walks in a large network with an irregular topology is infeasible because a central authority could not maintain accurate information about a dynamic set of members. The solution is to exploit the fact that sampling with a given probability distribution  can be simulated by a discrete-time Markov chain. Consider an irreducible Markov chain with states $(i,j) \in \{0, 1, \ldots ,S \}$ and let $P=[p_{ij}]$ denote its probability transition matrix where

\begin{equation}
p_{ij} = \text{Prob} [X(t+1) = j \mid X(t) = i]
\end{equation}
with $X(t)$ the state  at time $t$. Let $\pi=(\pi_{0},\pi_{1}, \ldots \pi_{S})$ be a probability distribution with nonzero probability for every state,  $\pi_{i} > 0,~0 \le i \le S$. The transition matrix $P$ is chosen so that $\pi$ is its unique stationary distribution thus, the reversibility condition $\pi = \pi P$ holds. When $g(.)$ is a function defined on the states of the Markov channel and we wish to estimate

\begin{equation}
E= \sum_{i=0}^{S} g(i) \pi_{i}
\end{equation}
we can simulate the Markov chain at times $t=1,2, \ldots, N$ and the quantity
\begin{equation}
\hat{E}= \sum_{i=1}^{N} {f(X(t)) \over N }
\end{equation}
is a good estimate of $E$, more precisely $\hat{E} \mapsto E ~\text{when}~ N \mapsto \infty$.
Hastings \cite{Hastings70} generalizes the sampling method of Metropolis \cite{Metropolis53} to construct the transition matrix given the distribution $\pi$. He starts by imposing the reversibility condition

\begin{equation}
\pi_{i} p_{ij} = \pi_{j} p_{ji}.
\end{equation}
If $Q=[q_{ij}]$ is the transition matrix of an arbitrary Markov chain on the states $\{0, 1, \ldots, S \}$ it is assumed that

\begin{equation}
p_{ij} = q_{ij} \alpha_{ij} ~~\text{if}~~ i \ne j~~~~\text{and}~~~p_{ii}=1 - \sum_{j \ne i} p_{ij}.
\end{equation}
Two versions of sampling are discussed in \cite{Hastings70}, the one of Metropolis and one proposed by Baker\cite{Baker65}; the quantities $\alpha_{ij}$ are respectively:

\begin{equation}
\alpha_{ij}^{M} =
\left\{
\begin{array}{lll}
1 & \text{if} & {\pi_{j} \over \pi_{i} } \ge 1 \\ \\
{\pi_{j} \over \pi_{i}} & \text{if} & {\pi_{j} \over \pi_{i}} < 1
\end{array}
\right.
\end{equation}

\begin{equation}
\alpha_{ij}^{B} = { \pi_{j} \over {\pi_i} + \pi_{j}}
\end{equation}
For example, consider a Poisson distribution $\pi_{i} = \lambda^{i} e^{-\lambda}/i!$; we  choose $q_{ij} = {1/2}$ if $j=i-1, i \ne 0$ or $j=i+1, i \ne 0$ and $q_{00} = q_{01}= 1/2$.
Then using Baker's approach we have

\begin{equation}
p_{ij} =
\left\{
\begin{array} {lll}
\lambda / ( \lambda + i + 1) & \text{if} & j=i+1, i \ne 0  \\
i/(i+\lambda) & \text{if} & j= i-1, i \ne 0 \\
\end{array}
\right.
\end{equation}
and $p_{00} = 1/2$ and $p_{01}= \lambda e^{-\lambda} / ( 1 + \lambda e^{-\lambda})$.

The algorithm to construct scale-free overlay topologies with an adjustable exponent in \cite{Scholtes10} adopts the equilibrium model discussed in \cite{Goh01}. The algorithm is based on random walks in a connected overlay network $G(V,E)$ viewed as a Markov chain with state space $V$ and a stationary distribution with a random walk bias configured according to a Metropolis-Hastings chain \cite{Hastings70}. Recall that in this case we assign a weight $p_{i}= i^{-\alpha},~ 1 \le i \le N, \alpha \in [0,1)$ to each vertex and add an edge between two vertices $a$ and $b$ with probability $p_{a} / \sum_{i=1}^{N} p_{i} \times p_{b} / \sum_{i=1}^{N} p_{i} $ if none exists; they repeat the process until $mN$ edges are created and the mean degree is $2m$. Then the degree distribution is

\begin{equation}
p(k) ~\sim k^{-\gamma}, ~~~~\text{with}~~~~ \gamma= (1+ \alpha) /\alpha.
\end{equation}

The elements of the transition matrix $P=[p_{ij}]$ are

\begin{equation}
\label{BiasEqun}
p_{ij}=
\left\{
\begin{array}{ll}
{1 \over k_{i} } \min \left\{  \left( {1 \over j} \right)^{1 \over {\gamma- 1}} {k_{i} \over k_{j}}, 1 \right\}
& (i,j) \in E \\
1 - { 1 \over k_{i} } \sum_{(l,i) \in E} p_{li} & i = j \\
0 & (i,j) \notin E
\end{array}
\right.
\end{equation}
with $k_{i}$ the degree of vertex $i$. An upper bound for the number of random walk steps can be determined from a lower bound for the second smallest eigenvalue of the transition matrix, a non-trivial problem.

Next we discuss how to construct the transition matrix $P$ when $\pi$ is a $d$-dimensional distribution; in this case the state at time $t$ is a vector with $d$ components $X(t)=(X_{1}(t), X_{2}(t), \ldots, X_{d}(t))$. Several techniques could be used to implement the state transition from $t$ to $t+1$:  (i) change all coordinates
of $X(t)$; (ii) randomly pick up one of the $d$ coordinates and change only the one selected; (iii) change only one coordinate at each transition but select the coordinates in a fixed order, rather than randomly.

The last techniques requires changing only one coordinate at each transition in the order $1, 2, 3, \ldots d$ and assumes that the process is observed only at times $0, 2d,3d, \ldots $. Call $P_{m}$ the transition matrix when only coordinate $m$ of $X(t)$ is affected.
Then the resulting process is a Markov process with the transition matrix $
P= P_{1} P_{2}, \ldots P_{d}$. Then $\pi$ is a stationary distribution when each $P_{m}$ satisfies the reversibility condition
$\pi P_{m} = \pi$.

%*********************************************************************
\subsection{An algorithm for the construction of scale-free networks}
\label{DistributedSFNAlgorithm}
%**********************************************************************

The algorithm to generate the scale-free network $\Gamma$ with $N$ nodes and $\mid E \mid$ edges assumes that each node has a unique ID, $nId$,  $1 \le nId \le N$.

The algorithm  consists of the following steps:

\begin{enumerate}
\item
Set $L$ the random walk length, e.g., $L=10$.
\item
Set the number of nodes already rewired, $n_{rewired}=0$.
\item
Select at random a node e.g., node {\bf a} and check if it has any edge that has not been rewired yet.
\begin{enumerate}
\item
If NO repeat step $3$.
\item
If YES pick up one of the edges at random and save both endpoints of that edge.
\end{enumerate}
\item
Check which one of the endpoints has higher degree, if they were same pick one of at random.
\item
Initialize the number of hops for the random walk $n_{hop}=0$.
\item
Draw a random number $0 < \kappa < 1$.
\item
Pick up at random a node in the neighborhood of the current node {\bf a}, e.g. node {\bf b}.
\item
Given the degree $d_{a}$ of node {\bf a} with  $vId_{a}$ and the degree $d_{b}$ of node {\bf b} with  $vId_{b}$ calculate
\begin{equation}
h = {d_{a} \over d_{b} } \left[ {nId_{a} \over nId_{b}} \right]^{1 \over {\alpha \gamma-1}}.
\end{equation}
\begin{enumerate}
\item
If  $h > \kappa$ choose node {\bf b}.
\item
If  $h \le \kappa$ choose node {\bf a}.
\end{enumerate}
\item
Increment the number of hops  $n_{hop}= n_{hop} +1$.
\begin{enumerate}
\item
If $n_{hop} \ne L$ and $n_{hop} < L$  go to Step 6.
\item
If  $n_{hop}=L$ save the node as the target node {\bf c}  then go to Step 6.
\item
Else save the node as the second target node {\bf d}.
\end{enumerate}
\item
Connect target nodes to each other.
\item
Remove the edge found in Step $3b$.
\item
Mark the edge you found as a rewired edge.
\item
Increment the number of nodes already rewired, $n_{rewired}= n_{rewired} +1$.
\begin{enumerate}
\item
If $n_{rewired} \le E$ go to Step $3$.
\item
Else, the algorithm terminates as we have rewired all edges.
\end{enumerate}
\end{enumerate}

%************************************************
\subsection{A distributed clustering algorithm}
\label{DistributedClusteringAlgorithm}
%************************************************

We assume that a scale-free network has been created and that due to the scale of the system there is no single site holding the information about the entire system. Instead, each individual node has the following information:

\begin{enumerate}
\item
The tuple $(nodeId, netAddr)$  giving the node identity  and its network address.
\item
The degree of the node, $nodeDeg$.
\item
The threshold for the selection of the core nodes e.g., $nodeDeg \ge 10$; thus, each node can determine if it is a core or a service node.
\item
The $neighborList$, the  list of pairs $(nodeId, netAddress)$, of all nodes directly connected to the node.
\item
The time $t_{0}$ when the clustering processing should start.
\end{enumerate}

The goal of the algorithm is to identify: (a) The service nodes connected to a core node; (b) The network connecting the core nodes. The algorithm uses two types of messages: (1) {\it Type 1 - cluster initiation},  messages sent by a core node to all its neighbors; {\it Type 2 - request to join a cluster}, message sent by a service node to the core node at the shortest distance. We shall use a modified version of an epidemic algorithm when a service or a core node $s_{i}$ re-sends an incoming message from the service node $s_{i,j}$ to all other service nodes in its connected nodes list,  but $s_{i,j}$.

The distributed algorithm is asynchronous and, in absence of global knowledge, about the system a service node should be able to determine if information provided by a core node is delayed due to a slower communication link and when it should proceed to making the decision which cluster it should join. All nodes share a time interval, $\tau_{end}$, which can be used to set up a timer; when the timeout occurs, all service nodes start the decision process leading to selection of the cluster a service node decides to join.

\bigskip

At the end of this asynchronous algorithm each core node would have built two tables:

\begin{itemize}
\item
The \underline{cluster table}, {\it clusterTab}, the list of service nodes which joined the cluster built around the core node. For each service node the list includes:
\begin{enumerate}
\item
The $(nodeId, netAddr)$ of the service node requesting to join the cluster.
\item
The distance to the service node, given by the hop count, $hopCount$.
\item
The path from the core to the service nodes, a list of nodes traversed by the {\it Type 1} message to reach the service node.
\end{enumerate}
\item
The \underline{core table}, {\it coreTab}, the list of core nodes directly connected to the node.
\end{itemize}

\begin{figure}[!ht]
\begin{center}
\includegraphics[width=5cm]{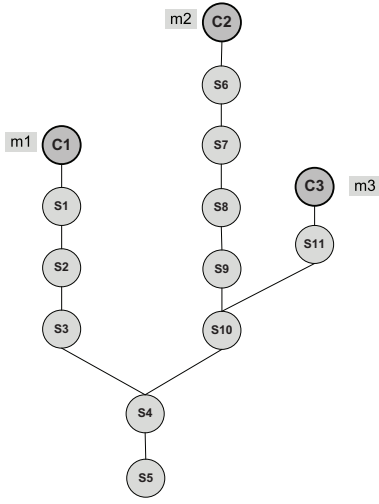}
\end{center}
\caption{The view of the world of service node $S4$; it is only aware of its neighbors, $S3, S5, S10$. The core nodes $(C1, C2, C3)$ send {\it Type 1} messages, $(m1, m2, m3)$, respectively. The distances of $S4$ to the three core nodes are: $d(S4, C3)=3, d(S4, C1)=4$, and $d(S4, C2)=6$. server node $S4$ will join the cluster of core node $C3$ at the minimum distance, $d=3$, if and only if message $m3$ arrives before $S4$ starts processing the information in its {\it tempTab}.}
\label{ExampleFig}
\end{figure}

Each service node maintains a table of all possible cluster it could join, the {\it tempTab}; at the end of the algorithm a service node will record the cluster it intends to join.

\medskip

At time $t_{0}$ all nodes start execution of the algorithm consisting of the following steps:

\begin{enumerate}
\item
All nodes start their  timer set to expire after $\tau_{end}$ units of time.
\item
Each core node sends to all nodes in its $neighborList$ a {\it Type 1} message containing its $(coreNodeId, netAddr)$ pair with a $hopCount=1$.
\item
Upon receiving a {\it Type 1} message,  a service node performs the following actions:
\begin{itemize}
\item
Parse the message and identify:
  \begin{enumerate}
  \item
  The core node sending the message, $(coreNodeId, netAddr)$;
  \item
  The path $msgPath$; and
  \item
  The number of hops, $hopCnt$.
  \end{enumerate}
\item
Check the {\it tempTab} for entries from the same core node.
\begin{enumerate}
\item
If no such entry exists:
 \begin{enumerate}
 \item
 Add to its $tempTab$ an entry consisting of:
 \begin{enumerate}
 \item
 Cluster Id, $clusterId=cordeNodeId$;
 \item
 The identity of neighbor delivering the message, $neigborId$;
 \item
 The  $hopCnt$;  and
 \item
 The path followed by the message, $msgPath$.
 \end{enumerate}
 \item
 Increment the $hopCnt$.
 \item
 Add itself to the $msgPath$.
 \item
 Resend the message to all neighbors, except the one which delivered the message.
 \end{enumerate}
\medskip
\item
If such an entry exists compare $hopCnt_{entry}$ of the existing entry with the one in the message, $hopCnt_{msg}$ .
  \begin{itemize}
  \item
  If $hopCnt_{msg} < hopCnt_{entry} $:
  \begin{enumerate}
  \item
  Replace the entry in the table with one containing the information in the message.
  \item
  Increment $hopCnt$.
  \item
  Add itself to the $msgPath$.
 \item
 Send the message to all neighbors except the one which delivered the message.
 \end{enumerate}
 \item
 If $hopCnt_{msg} \ge hopCnt_{entry} $  drop the message.
 \end{itemize}
 \end{enumerate}
 \end{itemize}
\item
When the timer $\tau_{end}$ expires a service node processes its {\it tempTab}. If there are no entries then the node is isolated and cannot join any cluster. \underline{Note:} service nodes of degree one can proceed with the actions discussed next once they get a Type 1 message; they do not need  to wait for the timer to expire because their commitment cannot be changed by any other message as their distance to the {\it core } node is one. All other service nodes should proceed as follows:
\begin{enumerate}
\item
Identifies the core node at the minimum distance.
\item
Retrieves from the entry:
\begin{enumerate}
\item
The identity of the core node,  $(coreNodeId, netAddr)$;
\item
The distance to the core node, $hopCnt$.
\item
The path to the core node, $msgPath$.
\end{enumerate}
\item
Constructs a {\it Type 2} message including this information.
\item
Sends the message to the core node.
\item
Resets the two timers.
\item
Stops
\end{enumerate}
\item
Upon receiving a {\it Type 2} message originating from  a service node, a core node:
\begin{enumerate}
\item
Processes the message to identify:
\begin{enumerate}
\item
The sender $(serviceNodeId, nodeAddr)$;
\item
The path to the service node, $msgPath$; and
\item
The distance to the service node, $hopCnt$.
\end{enumerate}
\item
Adda a new entry to its $clusterTab$
\end{enumerate}

\item
Upon receiving a {\it Type 1} message originating from  a core node, a core node:
\begin{enumerate}
\item
Retrieves from the entry, the identity of the core node,  $(coreNodeId, netAddr)$.
\item
Adds to its {\it coreTab} a new entry.
\end{enumerate}
\end{enumerate}

Once the clusters are constructed, a {\it service node} communicates only with the core node whose cluster it has joined; core nodes communicate with one another using an epidemic algorithm, each one forwards an incoming message to all its neighbors, except the one it has received the message from.

\medskip

The algorithm requires a timer because individual nodes do not have global information. Indeed, each node has only local information, it is aware of its neighbors and of its own degree. A service node does not know if a {\it Type 1} message from a node at a smaller distance from a core node was delayed and it will come after it has already received {\it Type 1} messages from all its neighbors.

This situation is illustrated in Figure \ref{ExampleFig} where we assume that the message $m3$ from core node $C3$ is delayed. We see that after receiving {\it Type 1} messages $m1$ and $m2$ from from both its neighbors, $S3$ and $S10$ then service node $S4$ would choose to join the cluster around core node $C1$ which is at distance 4 (core node $C2$ is at distance 5). On the other hand, if it waits for the  message $m3$ then service node $S4$ makes the correct decision. Thus, the time $\tau_{decision}$ should be chosen to ensure that all healthy nodes could transmit the {\it Type 1} messages they receive from core nodes.

%*********************************************************************
\subsection{Implementation and results}
\label{DistributedAlgorithmsResults}
%**********************************************************************

The distributed algorithms discussed in Sections \ref{DistributedSFNAlgorithm} and \ref{DistributedClusteringAlgorithm} were also implemented in Java. This time we used a much larger number of nodes $N=10^{5}$ and $N=10^{6}$, but maintained $\gamma=2.5$. The execution was done on the Amazon cloud using one {\it medium} instance. The execution times on the cloud are comparable with the times when execution was done locally on a system with similar resources as the ones provided by the AWS instance.

\begin{figure}[!ht]
\begin{center}
\includegraphics[width=7cm]{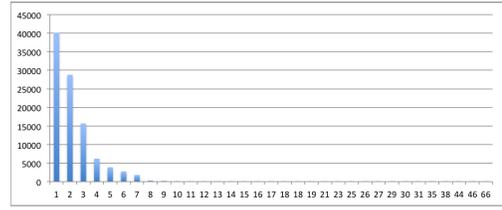}
\includegraphics[width=7cm]{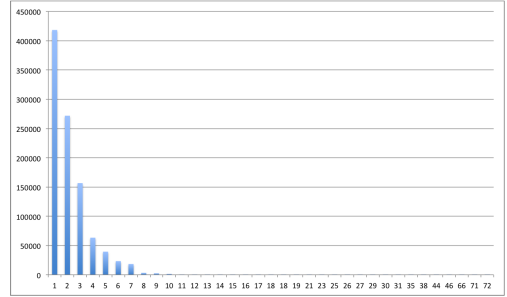}
\end{center}
\caption{The degree distribution of a scale-free network constructed with the algorithm in Section \ref{DistributedSFNAlgorithm}. (Top) $N=10^{5}$. (Bottom) $N=10^{6}$.}
\label{D-DegreeDistribution5Fig}
\end{figure}

\begin{figure}[!ht]
\begin{center}
\includegraphics[width=7cm]{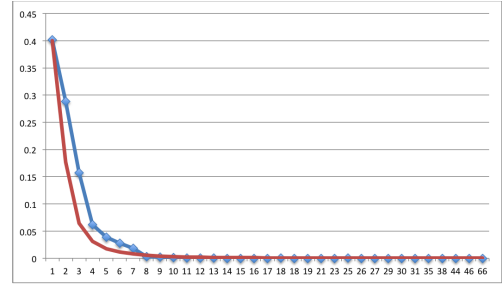}
\includegraphics[width=7cm]{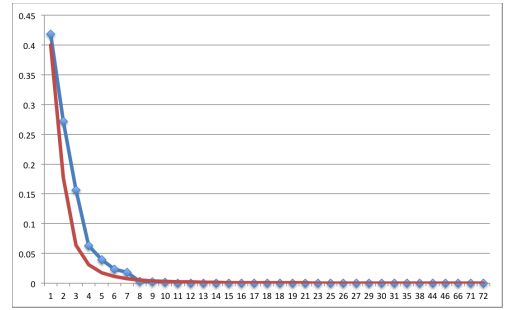}
\end{center}
\caption{The distributed algorithm to construct a scale-free network; theoretical distribution (red, continuous line) versus the degree distribution obtained with the algorithm introduces in Section \ref{DistributedAlgorithms} (blue line with dots). (Top) $N=10^{5}$ nodes; (Bottom) $N=10^{6}$ nodes.}
\label{D-ErrorFig}
\end{figure}

Figures \ref{D-DegreeDistribution5Fig} show the degree distribution of a scale-free network constructed with the algorithm in Section \ref{DistributedSFNAlgorithm} when $N=10^{5}$ and $N=10^{6}$ and Figures \ref{D-ErrorFig} plot the theoretical and the experimental degree distributions for the two cases.

Figure \ref{D-DegreeDistribution5Fig} shows a histogram of the degree distribution when the number of nodes is $N=10^{5}$. The distance between the theoretical and the experimental degree distribution is $e=0.2627$. The time for the construction of the scale-free network is $119$ seconds.

We use two thresholds for the separation of core nodes,  $T=10$ and $T=11$. Table \ref{ResultsTab} summarizes the number of core nodes thus, the number of clusters and the distance between the theoretical and the experimental degree distribution and the execution time for the two algorithms.

\begin{figure}[!ht]
\begin{center}
\includegraphics[width=7cm]{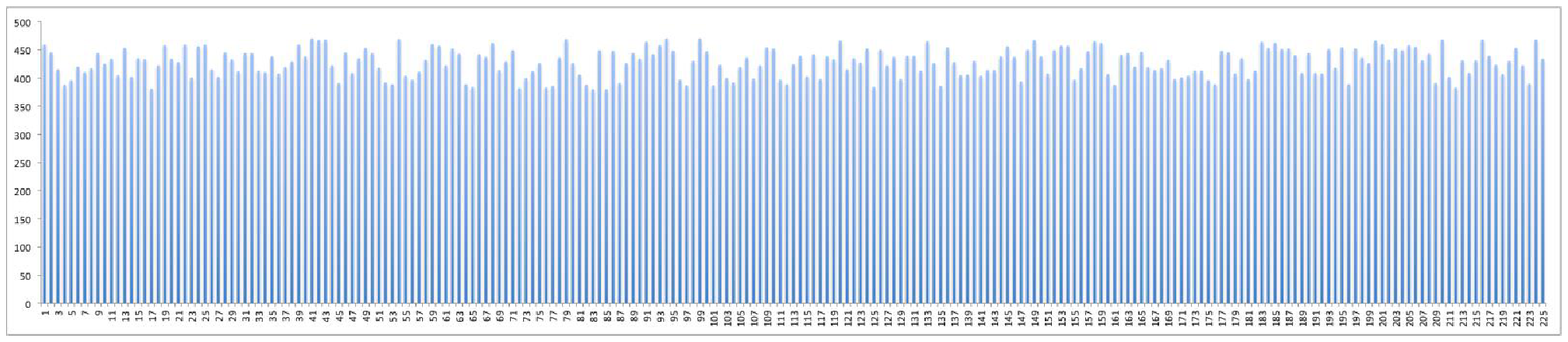}
\includegraphics[width=7cm]{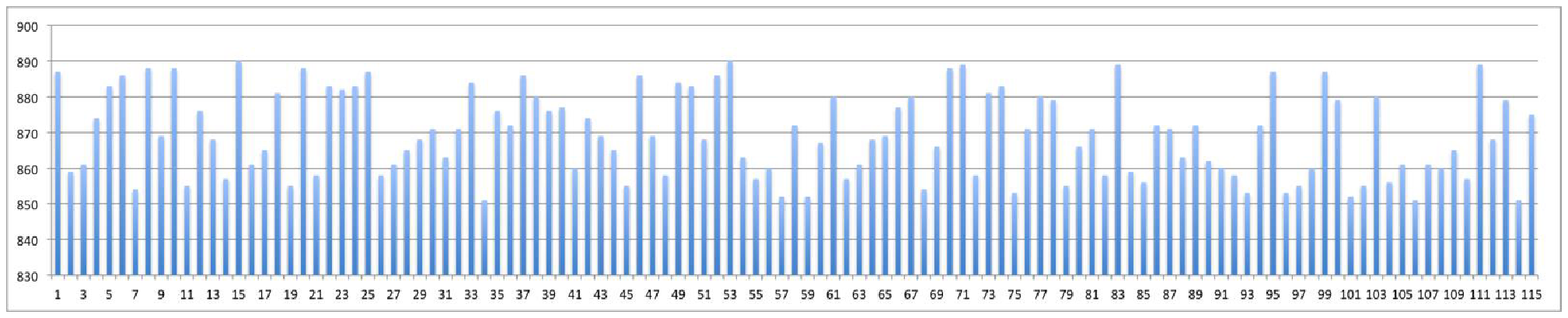}
\end{center}
\caption{The histogram of the cluster size w. (Top) $T=10$ and $M=225$.  (Bottom) $T=11$ and $M=115$.}
\label{D-ClusterSizeFig}
\end{figure}

Figures \ref{D-ClusterSizeFig} show the histogram of the the cluster size when the number of nodes is $N=10^{5}$ for $T=10$ and $T=11$. When $N=10^{6}$ due to the very large number of clusters the distribution of the cluster size $\mathcal{C}$ cannot be represented graphically. When $T=10$ the average cluster size is $\mu_{\mathcal{C}}=387$, the standard deviation is $\sigma_{\mathcal{C}}=43.74$ and the variance $\text{Var}_{\mathcal{C}}=1,913$. When $T=11$ the average cluster size is $\mu_{\mathcal{C}}=985$, the standard deviation is $\sigma_{\mathcal{C}}=46.1$ and the variance $\text{Var}_{\mathcal{C}}=2,125$.

\begin{table*}[ht!]
\begin{center}
\begin{tabular} {|c|c|c|c|c|c|c|}
\hline
 \# of nodes  &  Distance & Threshold  & Number of     & SFN time  & Clustering time & Total time \\
    (N)       &    (e) &   (T)      & clusters (M)  & (seconds) & (seconds)       &  (seconds) \\
 \hline
 $10^{5}$     & 0.2627 &  10 & 115   &  119  & 326   & 445 \\
              & 0.2627 &  11 & 225   &  119  & 476   & 596 \\
 \hline
 $10^{6}$     & 0.2578 &  10 & 985   & 329   & 793   &  1,022 \\
              & 0.2578 & 11 & 2,584 &  329  & 1,096 &  1,425 \\
 \hline
\end{tabular}
\caption{Summary of the results for the creation of a scale-free network with the algorithm in Section \ref{DistributedSFNAlgorithm} and for clustering using the algorithm in Section \ref{DistributedClusteringAlgorithm}. The distance between the theoretical and the experimental degree distribution and the execution time required by the algorithms.}
\label{ResultsTab}
\end{center}
\end{table*}

These results show that our algorithm achieves a relatively good approximation of the theoretical degree distribution, $0.25 < e < 0.265$ in a reasonably short time; the execution time of the algorithm increases as the logarithm of the number of nodes. On the other hand, clustering is more computationally intensive and the clustering time increases with the number of clusters and the with the number of nodes as shown in Table \ref{ResultsTab}.

\section{Application to Cloud Self-management}
\label{ComputerClouds}

A computer cloud is a complex system with a very large number of shared resources subject to unpredictable requests and affected by external events it cannot control. Cloud resource management is extremely challenging; the complexity of the system makes it impossible to have accurate global state information because of the interactions with the environment.

We believe that a scale-free  overlay network enables self-organization of a large-scale system, provides a natural way to select a distinct subset of nodes of the cloud computing infrastructure and build clusters of service nodes around them. The core nodes use only local information thus, are in a better position to efficiently implement optimal resource management policies.

%****************************************
\subsection{Cloud resource management}
\label{CloudResourceManagement}
\medskip
%***************************************

The policies for cloud resource management can be loosely grouped into five classes: (1) admission control; (2) capacity allocation; (3) load balancing; (4) energy optimization; and  (5) quality of service (QoS) guarantees.

The explicit goal of an admission control policy is to prevent the system from accepting workload in violation of high-level system policies \cite{Gupta09}; for example, a system may not accept additional  workload which would prevent it from completing work already in progress or contracted. Limiting the workload requires some knowledge of the global state of the system.  Capacity allocation means to allocate resources for individual instances; an instance is an activation of a service.   Locating  resources subject to multiple global optimization constraints requires a search in a very large search space when the state of individual systems changes rapidly. Load balancing and energy optimization are correlated and affect the cost of providing the services; they  can be done locally, but global load balancing and energy optimization policies encounter the same difficulties as the the capacity allocation \cite{Kusic08}. Quality of service is probably the most challenging aspect of resource management and, at the same time, possibly the most critical for the future of cloud computing.

The resource management policies must be based on a disciplined approach, rather than ad hoc methods. Basic mechanisms for the implementation of resource management policies are:

\smallskip

\noindent {\it Control theory.} Control theory uses the feedback to guarantee system stability and to predict transient behavior  \cite{Kusic08}, but can be used only to predict local, rather than global behavior;  applications of control theory to resource allocation are covered in \cite{Dutreilh10}. Kalman filters have been used for unrealistically simplified models as reported in \cite{Kalyvianaki09}, and the placement of application controllers is the topic of \cite{Tang10}.

\smallskip

\noindent {\it Machine learning.}  A major advantage of machine learning techniques is that they do not need a performance model of the system \cite{Tung07}; this technique could be applied for coordination of several autonomic system managers as discussed in \cite{Kephart07}.

\smallskip

\noindent {\it Utility-based.} Utility based approaches require a performance model and a mechanism to correlate user-level performance with cost \cite{Kephart11}.

\smallskip

\noindent {\it Economic models.} Auction models, such as the one discussed in \cite{Stokely10}, cost-utility models, or macroeconomic models are an intriguing alternative and have been the focus of research in recent years.

To our knowledge, none of the optimal, or near-optimal, methods to address the five classes of policies  scale up thus, there is a need to develop novel strategies for resource management in a computer cloud.  Typically, these methods target a single aspect of resource management, e.g., admission control, but ignore energy conservation; many require very complex computations that cannot be done effectively in the time available to respond. The performance models required by some of the methods are very complex, analytical solutions are intractable, and the monitoring systems used to gather state information for these models can be too intrusive and unable to provide accurate data. Many techniques are concentrated on system performance in terms of throughput and time in system, but they rarely include energy trade-offs or QoS guarantees. Some techniques are based on unrealistic assumptions; for example, capacity allocation is viewed as an optimization problem, but under the assumption that servers are protected from overload.

Virtually all mechanisms for the implementation of the resource management policies  require the presence of a few systems which monitor and control the entire cloud, while the large majority of systems run applications and store data;  some of these mechanisms require a two-level control, one at the cloud level and one at the application level. The strategies for resource management associated with IaaS (infrastructure as a service), PaaS (platform as a service), and SaaS (software as a service) will be different, but in all cases the providers are faced with large fluctuating loads. In some cases, when a spike can be predicted, the resources can be provisioned in advance, e.g., for Web services subject to seasonal spikes. For an unplanned spike the situation is slightly more complicated. Auto-scaling can be used for unplanned spike loads provided that: (a) there is a pool of resources that can be released or allocated on demand and (b) there is a monitoring system which allows a control loop to decide in real time to reallocate resources. Auto-scaling is supported by PaaS services, such as Google App Engine. Auto-scaling for IaaS is complicated due to the lack of standards; the OCCI (Open Cloud Computing Interface), an organization within OGF (Open Grid Forum) is involved in the definition of virtualization formats and APIs for IaaS.

A fair number of papers cover different facets of resource management in cloud computing, e.g., \cite{Aoun10} and \cite{ChangF10}. Scheduling of realtime services is discussed  in \cite{Liu10}. The performance management for cluster-based Web services is covered in \cite{Pacifici05}. Auctions in which participants can bid on combinations of items or {\it packages} are called {\it combinatorial auctions}; such auctions provide a relatively simple, scalable, and tractable solution to cloud resource allocation. Two recent combinatorial auction algorithms are the {\it Simultaneous Clock Auction} and the {\it Clock Proxy Auction}  \cite{Ausubel06}; the algorithm introduced in \cite{Stokely10}  is called {\it Ascending Clock Auction}.

%****************************************************
\subsection{Self-management and scale-free networks}
\label{CloudSelfManagement}
%****************************************************

In a cloud where changes are frequent and unpredictable,  centralized control is unlikely to provide continuous service and performance guarantees.Autonomic policies are of great interest in cloud computing due to the scale of the system, the large number of service requests, the large user population, and the unpredictability of the load; indeed, the ratio of the average to the peak resource needs of an application can be very large. Several papers are dedicated to the subject; for example, \cite{Ardagna11} covers energy-aware resource allocation, while \cite{Kephart11} analyzes policies based on utility functions for autonomic computing. Coordination of multiple autonomic managers and power-performance tradeoffs are presented in \cite{Kephart07}, while \cite{Addis10} analyzes autonomic management of cloud services subject to availability guarantees and \cite{Steinder08} covers autonomic management of heterogeneous workloads.

We start our analysis of cloud resource management with the question: What are the most desirable properties of a Resource Management System (RMS) for optimal capacity allocation, load balancing, energy optimization, and QoS guarantees in a computer cloud?

There is a wide agreement within the community that self-organization, self-management, and self-repair are highly desirable attributes of a computer cloud organization. Thus, the RMS should support the autonomic creation of two groups of cloud nodes with different roles: a small subset of cloud systems should assume control functions and initiate the creation of virtual clusters;  the others should assume the role of servers/workers. Individual systems should decide whether to join a virtual cluster and this decision should be based on local information and on the SLA requirements.

The RMS should facilitate the acquisition of {\it packages of resources}, or resource bundling, similarly to combinatorial auctions; indeed, an instance of an application needs a bundle of resources including CPU cycles, main memory, disk space, networking bandwidth. The mechanisms used by the RMS  should not require a model of the system and should be flexible and work well with the four classes of techniques discussed in Section \ref{CloudResourceManagement}. Optimal RMS policies require a monitoring system able to gather accurate state information with minimal system overhead. The most important requirement is that the mechanisms used by the RMS should support optimal policies for {\it all aspects of resource management} including capacity allocation, load balancing, energy optimization, and QoS guarantees.

The communication infrastructure is a critical component of a large-scale distributed system and overlay networks are ubiquitous in peer-to-peer systems and in other systems based on the client-server paradigm. An overlay network, or a virtual cloud interconnect, could be designed to respond to the requirements for the implementation of the resource management system; the overlay network should be scalable and it should facilitate an effective implementation of:

\smallskip

\noindent  (1) Algorithms for the selection of the nodes assigned control functions and for  efficient clustering of server nodes to control nodes; the network should enable a control node to gather accurate state information about its satellite server nodes by minimizing the average distance between them.

\smallskip

\noindent  (2) A variety of mechanisms for resource management policies. For example, facilitate the coordination of multiple autonomic controllers as discussed in \cite{Kephart07}, allow parallel auctions to be carried out extending the methods in \cite{Stokely10}, support the applications of control theoretical principles for resource management \cite{Kusic08}, support real-time applications \cite{Lin07}, and support the strategy based on random walks for selecting nodes with desirable properties discussed in this proposal.

\smallskip

Scale-free networks satisfy all these conditions; moreover, they allow server nodes to maintain a minimum of information about the network topology, they only need to know the network addresses of nodes they are connected to and the cluster they belong to. At the same time, core nodes need only be aware of the network addresses of the server nodes at distance one from them.

All communication from a core node to the server nodes in the cluster can be done by broadcasting to server nodes at distance one which are then required to forward all messages to the other server nodes they are connected to. A core node could request status reports from all servers in the cluster and can also distribute the workload assigned to the cluster based on more accurate state information than in other network configurations.

\section{Conclusions and Future Work}
\label{Conclusions}

System scalability should be an ab-initio concern in the design of any large-scale  system and in particular of a cloud computing infrastructure. A very large number of servers have to work in concert and they have to communicate effectively. The topology of the interconnection network which allows the servers  to communicate with one another is critical for ensuring system scalability and for creating the conditions for the implementation of optimal resource management policies. We argue that the constraints of a physical interconnection topology of a system can be overcome by designing an overlay network that enables scalability and allows the system to perform its functions in an optimal way.

Scale-free networks enjoy a set of desirable properties, they are non-homogeneous, resilient to congestion, robust against random failures, and have a small diameter and short average path length, as discussed in Section \ref{ScaleFreeNetworks}. The analysis  and the results presented in Sections \ref{CentralizedAlgorithm} and \ref{DistributedAlgorithms} show that efficient algorithms to construct such networks and to assemble clusters of servers around the core nodes of the scale-free network can be implemented with relative ease. Centralized algorithms can be used when the number of system components is relatively small, in the range of $10^{3}$, while distributed algorithms are useful when the number of components is several orders of magnitude larger, e.g., $10^{6}$ or larger. Distributed algorithms based on biased random walks support self-organization in a large-scale system; they allow us not only to construct scale-free networks, but also to select components with a set of desirable properties.

The algorithms for the construction of scale-free networks and for clustering discussed in this paper are particulary useful for the self-management in a computing cloud where individual core nodes of the global scale-free network could serve as cloud access points. Once such clusters are formed one can implement optimally the five classes of resource management policies; for example, each core node could request the creation of level-2 clusters in response to different requirements imposed by Service Level Agreements.  Level-2 clusters can be assembled through a random walk from the servers in a Level 1 cluster which satisfy security, location, QoS, and other types of constraints.

The biased random walk process discussed in Section \ref{BiasedRandomWalks} could be used to assemble the hybrid clouds with some servers in the private cloud of an organization while others are in public clouds. For example, a smart power grid application would require multiple electric utility companies to use a public cloud to trade and transfer energy from one to another, while maintaining confidential  information on servers securely located in their own private cloud. Similar configurations are likely for a unified health care system or any other application involving multiple organizations required to cooperate with one another but with strict privacy concerns.

The solutions we propose represent a major departure from the organization of existing computing clouds. The significant advantages of self-organization and self-management, in particular the ability to implement effective admission control and QoS policies reflecting stricter SLA requirements, seem important enough to justify the need for a paradigm shift in cloud computing organization and management. Indeed, over-provisioning used by exiting clouds is not a sustainable  strategy to guarantee QoS. It seems reasonable to expect that in the future the providers of Infrastructure as a Service (IaaS) cloud delivery model will support applications with strict security and response time constraints, while lowering the cost of services through better resource utilization; we believe that this can only be done in a self-organizing system and a scale-free virtual interconnection network seems ideal for such a system.

Our future work will be focused on the development of communication algorithms among the core nodes, an in-depth investigation of the fault-tolerance of the systems we propose, and on the study of systems ability to respond to sudden load surges. We plan to create a test-bed system to compare the implementation of different resource management policies in a system based on self-organization principles and one with a hierarchical  control structure. A fair and unbiased comparison of resource management policies and their implementation seems to be a rather non-trivial task.

\bigskip

{\it Ashkan Paya.} Ashkan Paya is a second year graduate student in the EECS Department at University of Central Florida pursuing a Ph.D. degree in Computer Science. He graduated from Sharif University of Technology in Teheran, Iran, with a BS Degree in Computer Science in 2011. His research interests are in the area of resource management in large-scale systems and in cloud computing

\bigskip

{\it Dan C. Marinescu.} In 1984 Dan Marinescu joined the Computer Science Department at Purdue University in West Lafayette, Indiana as an Associate and the Full Professor. Since August 2001 he is a Provost Research Professor and Professor of Computer Science at University of Central Florida. His research interests are: scientific computing, process coordination and distributed computing including cloud computing and quantum information processing. He has published more than 210 papers in referred journals and conference proceedings. He published several books: {\it Internet-based Workflow Management} published by Wiley in 2002, {\it Approaching Quantum Computing} (co-authored with Gabriela M. Marinescu, Prentice Hall - 2005); {\it Classical and Quantum Information}, (co-authored with Gabriela M. Marinescu) published in February 2011 by Academic Press, a division of Elsevier, and {\it Cloud Computing: Theory and Practice} published by Morgan Kaufmann in 2013.

\end{document}